\documentclass[aps,pre,showpacs,showkeys,12pt]{revtex4-1}
\usepackage[utf8x]{inputenc}
\usepackage{graphicx}
\usepackage{default}
\newcommand{\be}{\begin{equation}}
\newcommand{\bea}{\begin{eqnarray}}
\newcommand{\bc}{\begin{center}}            
\newcommand{\ee}{\end{equation}}
\newcommand{\eea}{\end{eqnarray}}
\newcommand{\ec}{\end{center}}
\newcommand{\baa}{\begin{eqnarray*}}
\newcommand{\eaa}{\end{eqnarray*}}
\begin{document}
\baselineskip 24pt
\title{On the form of prior for constrained thermodynamic processes with uncertainty}
\author{Preety Aneja and Ramandeep S. Johal}
\affiliation{Department of Physical Sciences,\\ 
Indian Institute of Science Education and Research Mohali, \\
Sector 81, Knowledge City, Manauli P.O., Ajit Garh -140306, India}
\begin{abstract}
 We consider the standard thermodynamic processes with constraints, but with
 additional uncertainty about the control parameters.  
 Motivated by inductive reasoning, we assign prior
 distribution that provides a rational
 guess about likely values of the uncertain parameters.
 The priors are derived explicitly for both the entropy conserving 
 and the energy conserving processes.  The proposed form is useful 
 when the constraint equation cannot be treated analytically.
 The inference is performed using spin-1/2 systems as models for heat reservoirs.
 Analytical results are derived in the high temperatures limit.
 Comparisons are found between the estimates of thermal quantities
 and the optimal values described by extremum principles.
 We also seek a intuitive interpretation of the prior
 and show that it becomes uniform over the quantity which is conserved 
 in the process. We find further points of correspondence
 between the inference based approach and the thermodynamic framework. 
\end{abstract}
\maketitle
\section{Introduction}
An objective choice for a prior consistent with the 
given prior information remained a long-standing problem in
Bayesian analysis \cite{Bernoulli2006, Laplace1774, Bayes1763,
Jeffreys1939,Jaynes1968,Agostini1999}. There have been different proposals for 
incorporating prior information via priors \cite{Wasserman1996, Press2003}. 
Ideally, ``prior information'' implies a piece
of knowledge about the system,
before any experimental data is considered. But it is seldom that 
we do not know anything beforehand, say, about a parameter for which 
we wish to assign a prior. It might be evident from the physical
laws or the constraints governing the model/system,
that the parameter is positive or restricted to 
a finite range. 
The prior that we seek in the absence of data, may, in a sense,
be regarded as minimally informative.

Such a prior is usually expected to
follow invariance under reparametrizations. Thus if
we are assigning prior $\pi$ for a (scale) parameter $\theta$, and we
are completely ignorant about the relevant scale in the system,
then a reparametrization like rescaling should not change
our state of knowledge \cite{Jeffreys1939, Jaynes1968}. 
More precisely, we impose: $\pi (\theta') d \theta' = \pi (\theta) d \theta$.
Then if $\theta' = c \theta$, where $c$ is a positive constant, it follows
that the prior must satisfy: $\pi (\theta) = c \pi (c \theta)$.
In this spirit, Jeffreys proposed his prior in the form, 
$\pi (\theta) = \sqrt{I(\theta)}$, where $I(\theta)$ is the Fisher information.
Actually, this form guarantees invariance of 
the prior under all continuous one-to-one
transformations. Anyway, the calculation of the function $I$ still requires the knowledge
of a model $f(x|\theta)$, which the data supposedly follow, given
a value of $\theta$. In this sense,  
Jeffreys' prior (and other proposals 
\cite{Skilling1989, Rodriguez1989, Zellner1991, Caticha2004} making use of
the likelihood $f(x|\theta)$) is not minimally informative.

On the other hand, one may question whether this 
invariance under all such transformations is really required \cite{Rao1987}. Rather, it seems
justified to impose this condition only for a class or subset of transformations, 
suggested by the particular problem under consideration.
Consider an example relevant to the theme of this paper, where
a system is composed of two similar subsystems. 
Each one is described by a property $T$, so we may label them
with values $T_1$ and $T_2$ \cite{Comment_random}. Suppose due to 
some constraint, these values 
satisfy a one-to-one relation: $T_1 = F(T_2)$. 
Clearly, the exact knowledge about $T_1$  yields a unique value $T_2$.
By the same token, an uncertainty in $T_1$, would imply a lack
of knowledge about $T_2$ also. 
Further, if the prior information does not  distinguish
between the labels, then it is reasonable
to treat our state of knowledge about either parameter as equivalent. 
The desideratum of consistency  
requires that we should assign the same form of prior distribution
to each of them \cite{Jaynes1968}. In other words, the prior  
should be invariant under a change of variable from $T_1$ to $T_2$.
Thus, here the invariance is demanded only for
a restricted class of transformations. 

In previous works, we considered the classic  problem of 
maximum work extraction from two finite reservoirs of heat \cite{PRD2012, PRD2013}.
In this process, the constraint of entropy conservation specifies a relation
between the two temperatures labeled $T_1$ and  $T_2$. 
But if we assume an ignorance 
about the exact value of $T_1$ or $T_2$, then our goal is
to make a rational guess about their likely values. 
From a Bayesian perspective, we seek 
a prior that reasonably quantifies our uncertainty of these variables.
Earlier \cite{PRD2013}, we derived priors using specific models of reservoirs
which yielded an explicit form of the relation  $T_1=F(T_2)$. 
The estimates of extracted work and the efficiency
at maximum expected work showed remarkable agreement with the optimal
features for these quantities. In particular, near equilibrium, 
a universal behavior for efficiency
is found to scale as  $\eta_c/2 + \eta_c ^2/8 + \cdots$, where 
$\eta_c$ is Carnot limit, and this feature could be  
inferred within the prior based approach also. 

In this paper, we cast the prior in
a general form and seek a more intuitive meaning 
for it within the standard thermodynamic framework. 
The derived prior can be applied to analogous processes 
where an explicit form of function $F(\cdot)$
is not feasible, a situation often
realised in physical systems. As a concrete example,
we study the case of spin-1/2 systems as our heat reservoirs. 
Apart from an entropy conserving process, we 
 apply the formalism to a pure thermal contact,
which is an energy conserving process.
Further, we provide interpretation  
of the form of prior and of the 
estimates in the context of thermodynamic framework.

The paper is organized as follows. In Section II, 
we derive the general form of prior for 
an entropy conserving process. In Section III, 
we present the model of reservoirs as $N$ spin-1/2 systems
and outline the procedure to conduct inference.
In succeeding Subsections, analytical formulae in
high temperature limit are derived and work, efficiency
of the process are estimated.
In Section IV, we apply the same approach to energy
conserving process between two reservoirs. 
Finally, Section V is devoted to discussing the 
meaning of priors and some concluding remarks.
\section{Assignment of prior}
To assign an appropriate prior, we first clearly state, 
the prior information about the system, and the assumptions
involved:

(i) the state of knowledge of an
observer is same irrespective of whether the   
uncertainty is quantified in terms of $T_1$ or $T_2$.
This would be plausible if each parameter is defined
in the same interval and the parameters are similar 
in nature (each represents
temperature). This notion is quantified by
assigning the {\it same} form of prior $P(T_i)$ to both parameters.
For convenience, we imagine two observers, each of which 
quantifies the uncertainty in terms of a specific temperature.

(ii) Each observer assigns the same probabilities
for the values of $T_1$ and $T_2$, constrained by the given process.
It implies
\be
P(T_2) = P(T_1) \left|\frac{dT_1}{dT_2}\right|.
\label{eqprob}
\ee
In certain cases \cite{PRD2013}, one may obtain an explicit function $F(\cdot)$ relating
$T_1$ and $T_2$.
Note that probabilities are being interpreted here in the sense
of degree of belief \cite{Baeirlein}.
The next step then is to solve for the function $P$ \cite{Comment_uniform}.
In the following, we will deduce consequences of this choice of the prior and 
make comparison between the estimates derived from the uniform prior and
the optimal characteristics of the process.

iii) The last bit of prior information is that the work is
extracted in the physical set up: $W = -\Delta U = U_+ + U_- -U_1 -U_2 \ge 0$.
As discussed in Ref. \cite{PRD2013}, when the  
reservoirs are identical except for their temperatures,
the work expression is invariant under the 
change of the labels for $T_1$ and $T_2$. 
Moroever, due to the constraint between 
these two variables,   
work may be regarded a function of one variable only.
Then using  the condition $W \ge 0$, we reuqire that  an uncertain temperature
can take values in the interval $[T_-,T_+]$.

It is apparent from Eq. (\ref{eqprob}), that a dependence between $T_1$ and $T_2$, 
should determine the form of prior. In particular,
we should know the rate of change of say $T_2$ with respect to $T_1$.

On the other hand, we may know a particular constraint governing the process.
For example, an entropy conserving process requires $dS = 0$, where 
$S$ is the total entropy of the reservoirs. Due to additive property
of entropy, we can write 
\be
dS_1 + dS_2 = 0,
\label{ds}
\ee
and further as:
\bea
\left(\frac{\partial S_1}{\partial U_1} \right)
\left(\frac{\partial U_1}{\partial T_1}\right) dT_1 +
\left(\frac{\partial S_2}{\partial U_2}\right)
\left(\frac{\partial U_2}{\partial T_2}\right) dT_2 & = & 0.
\eea
We assume that no work is performed on or by the heat reservoirs.
Using the definition of temperature, $({\partial S}/{\partial U})_V = {1}/{T}$ and 
heat capacity ${({\partial U}/{\partial T})}_V = C(T)$ in the above equation, we get:
\bea
\frac{dT_1}{dT_2} & = & -\frac{C_2/T_2}{C_1/T_1}.
\label{inf}
\eea
The above equation relates a infinitesimal change in one of the temperatures
to a corresponding change in the other temperature.
The negative sign indicates the opposite sign of the changes as the process 
advances. The ratio above on the lhs, if we interpret it as a rate of change,
is suggested by the constriant on the physical process, and forms 
a part of the prior information. So now we are
going to identify it with the rate of change as appearing in Eq. (\ref{eqprob}).
For the purpose
of the prior, we need only the magnitude of this
relative change: $|{dT_1}/{dT_2}| = {(C_2/T_2)}/{(C_1/T_1)}$.
So substituing in Eq. (\ref{eqprob}), we obtain
\be 
\frac{P(T_2)}{P(T_1)} = \frac{C_2/T_2}{C_1/T_1},
\ee
and by applying a separation of the variables, we can write:
\be
P(T_i)=\frac{C_i(T_i)/T_i}{N},
\label{pr}
\ee
where $i=1,2$ and $N=\int C_i(T_i)/T_i\;dT_i$ can be determined from the normalisation
condition on the prior.

In Ref. \cite{PRD2013}, the reservoirs were assumed to obey the fundamental thermodynamic relation:
$S\propto U^{\omega_1}$, where  
$\omega_1$ is a known constant. This implies 
$U \propto T^{{1}/{(1-\omega_1)}}$ and 
$C(T) \propto T^{\omega}$, where $\omega = \omega_1/(1-\omega_1)$. Considering such 
reservoirs for the concomitant process, the prior has the form:  
\be
P(T)=\frac{\omega {T}^{\omega-1}}{({T_+}^{\omega}-{T_-}^{\omega})}.
\label{l1}
\ee
The above prior was derived from the use of the integral
form of Eq. (\ref{ds}), given by $S_+ + S_- = S_1 + S_2$, alongwith Eq. (\ref{eqprob}).
We have derived above a general prior, which  is
also consistent with the special form of Eq. (\ref{l1}).
The general form is useful, in particular when we cannot 
write an explicit function $F(\cdot)$ that relates $T_1$ and $T_2$. 

Using Eq. (\ref{pr}), the estimate for $T_i$, defined as its average value, is given as:
\be
\overline{T}_i = \int_{T_-}^{T_+} T_i P(T_i)\; dT_i.
\label{aT2}
\ee
 However, 
the estimate by an observer (say 2) for the temperature of the other reservoir
is $\tilde{T}_1 = F(\overline{T}_2)$. After knowing the  estimates
for final temperatures, one can estimate other thermal
quantities, like the maximum work extracted and efficiency of the process. 
When it is not be possible to ascertain the functional 
form $F(\cdot)$, then after calculating $\overline{T}_2$,
the estimate for $\tilde{T}_1$ has to be performed numerically. 

\section{Model}{\label{m}}
As an application, we consider two {\it finite}, heat reservoirs at temperatures $T_+$ and $T_-$, 
each consisting of $N$ non-interacting, localized spin-1/2 particles.
A spin-1/2 particle can be regarded as a two-level system,  
with energy levels ($0$, $a$).   
The mean energy for such a reservoir is given by:
\be
U = \frac{Na e^{-{a}/{kT}}}{1+e^{-{a}/{kT}}},
\label{int}
\ee
where $k$ is Boltzmann's constant.
The heat capacity is given by:
\be
C = Nk \left(\frac{a}{kT}\right)^2 \frac{e^{-{a}/{kT}}}{(1+e^{-{a}/{kT}})^2}.
\label{hc}
\ee
The entropy for each reservoir can be written as:
\be
S = N \left[\ln {(1+e^{-{a}/{k T}})} + \frac{a}{k T} \frac{e^{-{a}/{k T}}}{1+e^{-{a}/{k T}}}\right],
\label{ent}
\ee 
Now using these finite reservoirs as the heat source and the sink respectively,
we consider the process of maximum work extraction by coupling them to an 
ideal engine. In this process, an infinitesimal 
amount of heat is extracted from the hot reservoir, converted into
work and the rest amount of infinitesimal heat is rejected to 
the cold reservoir.
The process stops when the reservoirs reach a
final common temperature, determined by the entropy conservation
condition.

Now consider that the process is  not yet completed and is at some intermediate
stage, given by temperatures $T_1$ and $T_2$.
In general, there is no explicit  
relation between $T_1$ and $T_2$, so that given a value for 
one temperature, the value for the other has to be determined
numerically. 
As will be shown below, the equation of entropy conservation 
can be solved in a closed form for $T_1$ in terms of $T_2$,
in the limit of high temperatures or when parameter $a$ is 
quite small compared to the reservoir temperatures. 

Now we  summarise the main steps in the estimation procedure.\\
1) Assign a prior for the uncertain temperatures.\\
2) Due to the constraint of entropy conservation, various quantities such 
as extracted work, are a function of one of the temperatures only. \\
3) The maximum work extracted is estimated by using the estimate for 
temperature, by substituting $ W(T_i) = W(\overline{T}_i)$, where 
$\overline{T}_i$ is the average based on the prior \cite{Comment_work}.\\
4) In contrast to the expression for work, 
the expressions for heat exchanges by the reservoirs are not symmetric w.r.t 
the two temperatures.
This implies that for calculations
on the efficiency of the process, the specific hot and cold reservoirs have
to be identified. 

For convenience, we choose in the following, $T_+ = 1$, $T_- = \theta$ and $k = 1$.
\subsection{High-Temperature Limit}{\label{highT}}
In case of very high temperatures as compared to
the level spacing $a$ i.e. $a \ll T$, we can solve the constraint equation 
analytically. Thus
keeping terms only upto $(a/T)^2$, we can write the relevant expressions as:
\bea
U &\approx& N \left[\frac{a}{2}-\frac{a^2}{4T}\right],
\label{int2}
\\
S &\approx& N \left[\ln 2 - \frac{a^2}{8T^2}\right],
\label{ent2} \\
C &\approx& N\left[\frac{a^2}{4 T^2}\right].
\label{hc2}
\eea
To find the relation between $T_1$ and $T_2$, we apply 
 $S_1+S_2 = S_+ + S_-$, and obtain:
\be
T_1  =  \frac{1}{\sqrt{1+\frac{1}{\theta^2}-\frac{1}{{T_2}^2}}}.
\label{rel}
\ee
Using Eqs. (\ref{hc2}), (\ref{pr}) in (\ref{aT2}), 
 we can estimate one of the temperatures ($T_2$) as:
\be
\overline{T}_2 = \frac{2 \theta}{1 + \theta}.
\label{av}
\ee
Then the other temperature $T_1$ is estimated from (\ref{rel}), 
just by substituting  $T_2 = \overline{T}_2$,
yielding $\tilde{T}_1  = 2 \theta / (\sqrt{3 \theta^2 -2 \theta+3})$. 
For a comparative study, we also consider 
the uniform prior over the range $[\theta, 1]$, which gives $\overline{T}_2 = (1+\theta)/2$
and so $\tilde{T}_1  = \theta (1+\theta)/ (\sqrt{(1+\theta^2)(1+\theta)^2-4\theta^2})$.
%
\subsection{Estimation of Work}
Work is defined as the difference of total initial and final energy of the reservoirs:
$W = U_+ + U_- - U_1 - U_2$. We know work can 
also be written as $W = Q_{\rm h} - Q_{\rm c}$.
Here we have departure from the standard thermodynamic
solution. There it is assumed a priori that one label say 1,
refers to initially hot reservoir and so the second label 2,
refers to the other reservoir. But we do not make any such 
assumption or include it in the prior information. 
Thereby just from the work expression in terms
of difference of initial and final energies,
we cannot assert a unique way in which $Q_{\rm h}$
or $Q_{\rm c}$ can be defined \cite{Comment_label}. This point will
be taken up again in the estimation of efficiency.

Now, by using Eq. (\ref{int2}), we obtain:
\be
W  =  \frac{Na^2}{4}\left(\frac{1}{T_1}+\frac{1}{T_2}-\frac{(1+\theta)}{\theta}\right).
\label{w2}
\ee
In terms of a single variable, using (\ref{rel}) we have
\be 
W(T_2)  =  \frac{Na^2}{4}\left( \sqrt{1+\frac{1}{\theta^2}-\frac{1}{{T_2}^2}}
+\frac{1}{T_2}-\frac{(1+\theta)}{\theta}\right).
\label{w2r}
\ee
At the optimality condition, $T_1=T_2=T_c$, 
\bea
T_c = \theta \sqrt{\frac{2}{1+\theta^2}}.
\label{topt}
\eea
So the optimal value of work $W_o$ is given by:
\bea
W_o = \frac{ N a^2}{4 \theta}\left[\sqrt{2 (1+\theta^2)}-(1+\theta)\right].
\eea
The extracted work is estimated by substituting the expected value for $T_2$ 
in Eq. (\ref{w2r}). Thus using (\ref{av}), we obtain:
\be
\tilde{W}_p  = \frac{N a^2}{8 \theta}\left[ \sqrt{3 \theta^2 -2 \theta+3} -(1+\theta)\right].
\label{w2es}
\ee
For the choice of a uniform prior, Eq. (\ref{w2r}) yields:
\be
\tilde{W}_u = \frac{N a^2}{4 \theta(1+\theta)}\left[ \sqrt{(1+\theta^2)(1+\theta)^2-4\theta^2}
      -(1+\theta^2)\right].
\label{wues}
      \ee
\begin{figure}[h]
\includegraphics[width=2.5in]{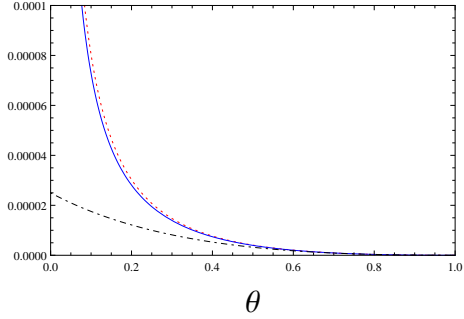} 
\caption{Work, scaled by $N$, as a function of $\theta$ for $a=0.01$. 
The dotted top curve is for optimal work $W_o$, 
middle solid curve is the estimate $\tilde{W}_p$, with the derived prior,
and the lower dotdashed curve is  
the estimate $\tilde{W}_u$, with uniform prior. 
The three curves agree in the near equilibrium
regime ($\theta \approx 1$).}
\end{figure}
      Note that the above estimates for work are the same for both the observers,
again due to symmetry in the work expression (\ref{w2}) w.r.t
$T_1$ and $T_2$.

Fig. 1 illustrates the comparison for a given value of $a$. The agreement 
between different estimates in the near-equilibrium regime ($\theta\approx 1$), 
can be studied by expanding the work estimates about $\theta =1$:
\be
\tilde{W}_p \approx \tilde{ W}_u = \frac{N a^2}{16}(1-\theta)^2 + 
\frac{3 N a^2}{32}(1-\theta)^3  + ~ O[1-\theta]^4.
\ee 
These estimates of work agree with the optimal work upto third order of $(1-\theta)$.
Fig. 2 shows the comparison of ${W}_o$, $\tilde{W}_p$ and $\tilde{W}_u$ for more general
values of parameter $a$, when the constraint of entropy conservation 
can be treated numerically only. We observe the close agreement in
the estimates for near-equilibrium. However, in general, the estimates
from the derived prior are much better estimates of the optimal work
than obtained from uniform prior, thus signifying the use of 
prior information in the assignment of the prior. 
\begin{figure}[h]
\begin{tabular}{ll}
\includegraphics[width=1.5in]{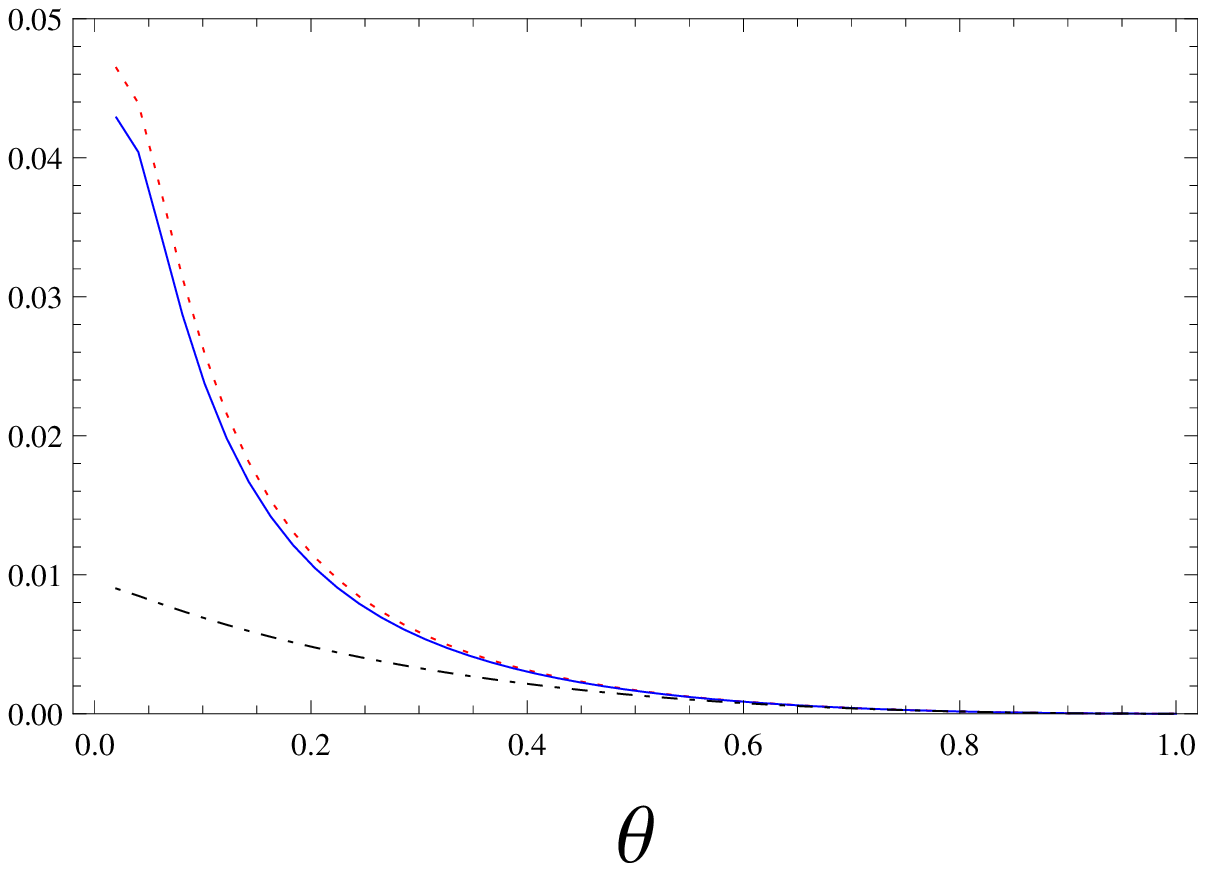} & \includegraphics[width=1.5in]{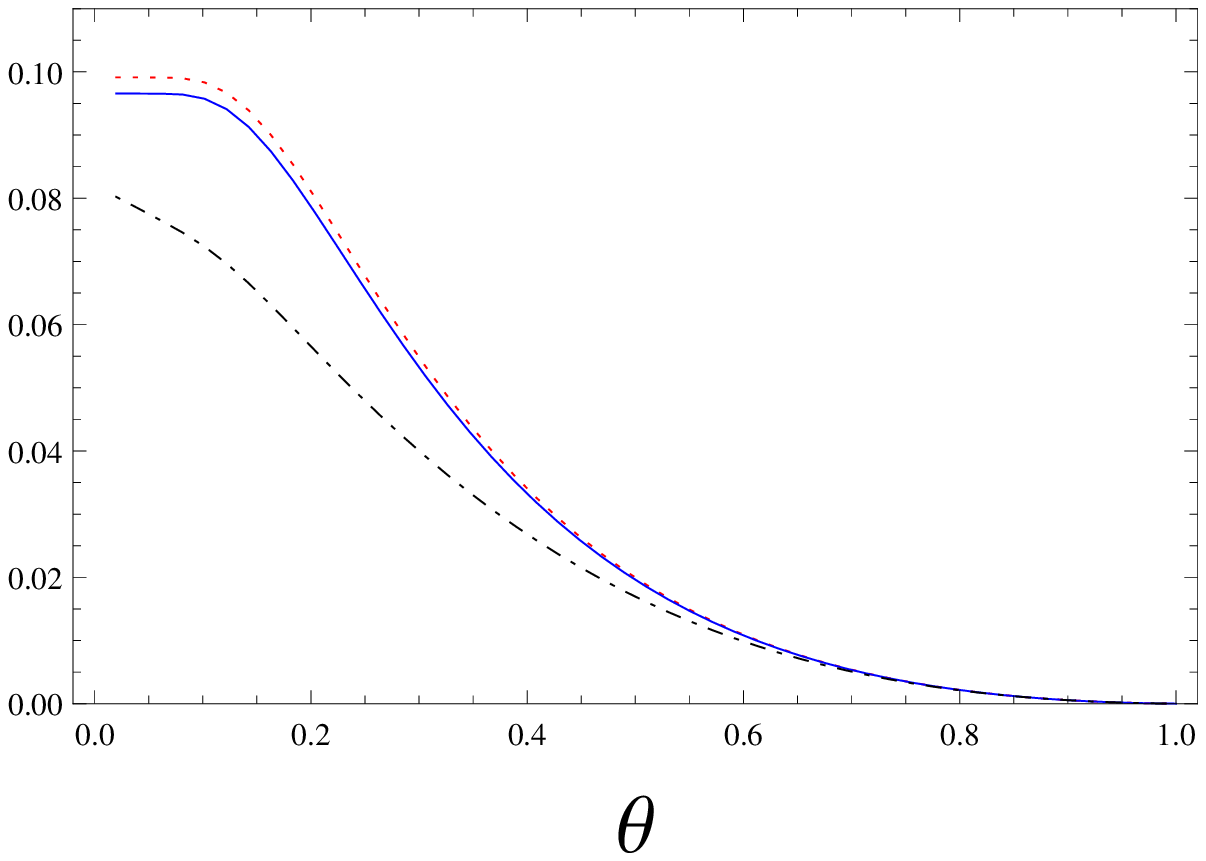}\\
\hspace{.7in}\small{\textit{(a)}} &  \hspace{.75in}\small{\textit{(b)}} \\\\
\includegraphics[width=1.5in]{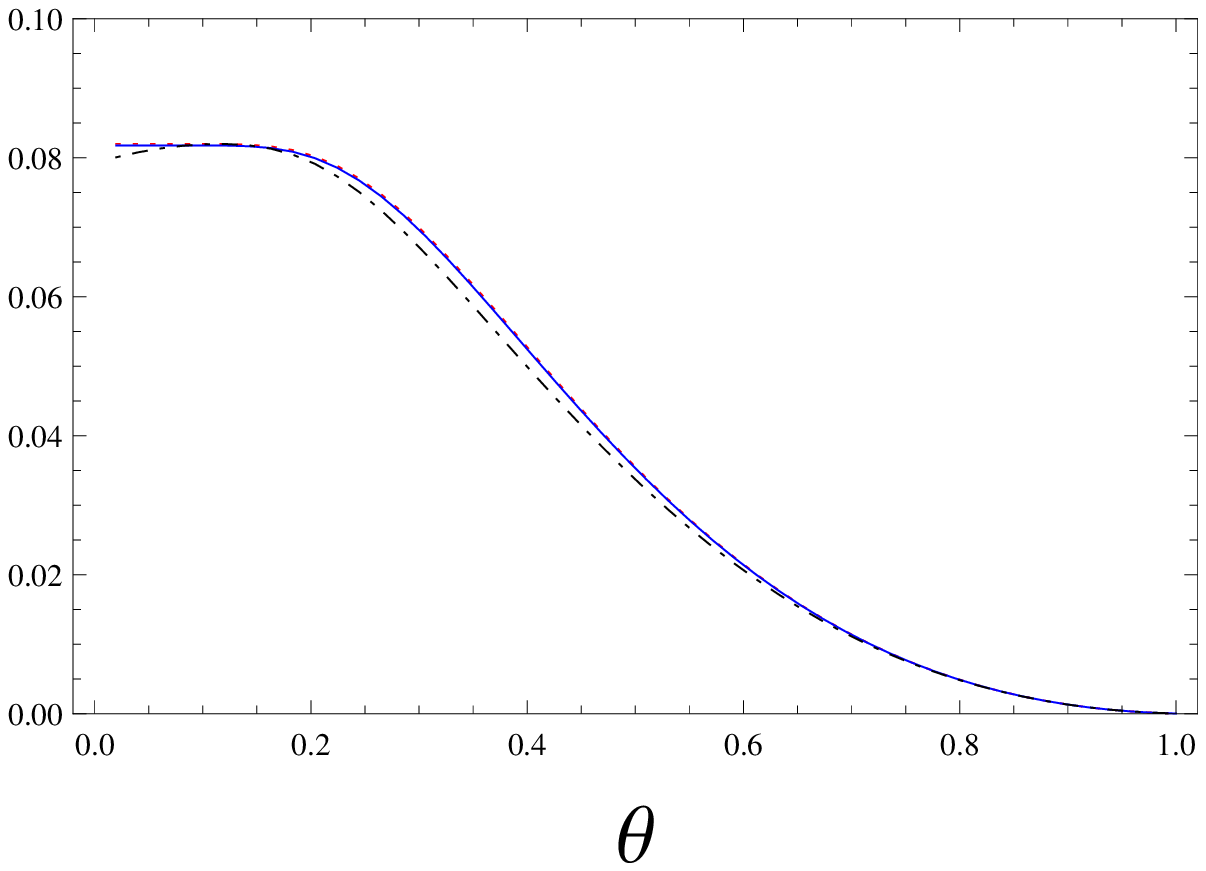}&\includegraphics[width=1.5in]{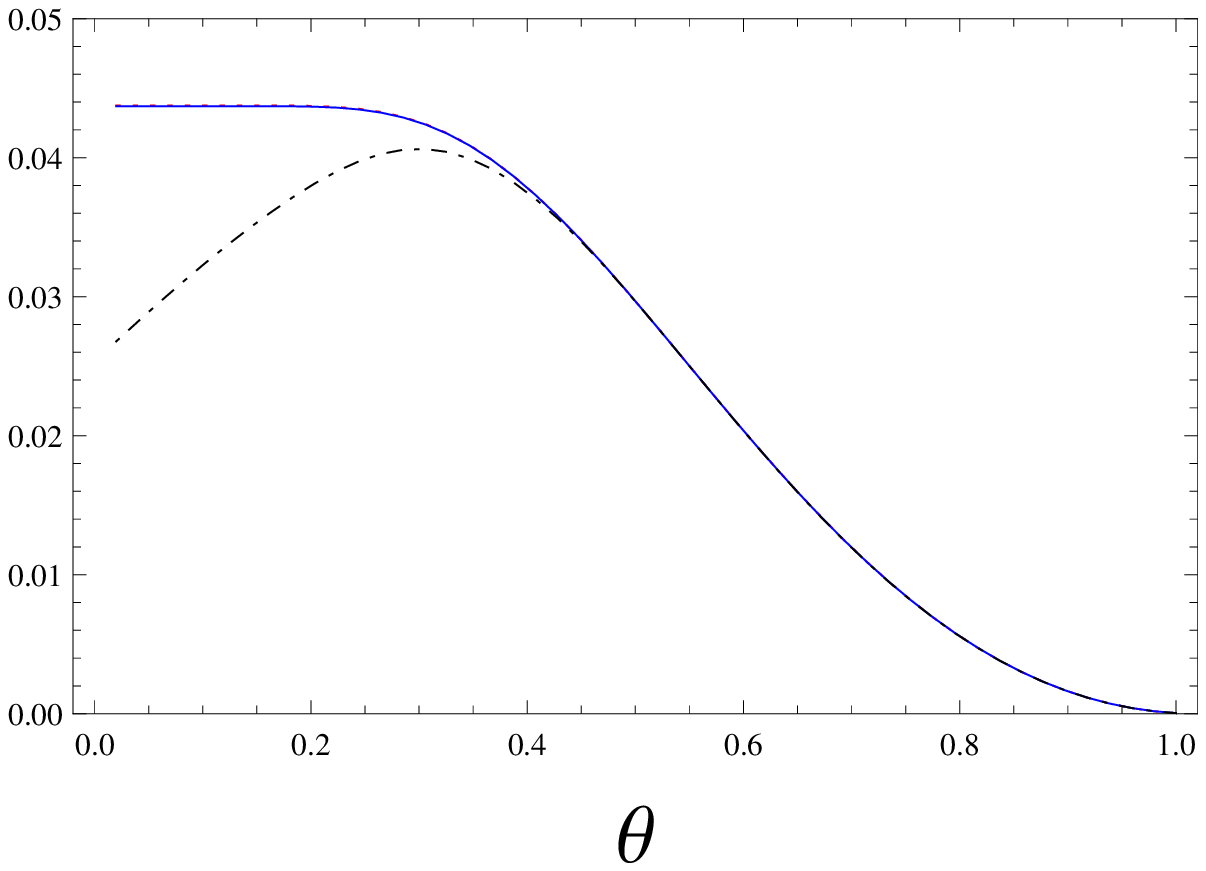}\\
\hspace{.75in}\small{\textit{(c)}} & \hspace{.75in}\small{\textit{(d)}}
\end{tabular}
\caption{Work, scaled by $N$, as a function of $\theta$ for different $a$'s 
values; (a) $a$ = 0.2,
(b) $a$ = 0.8 , (c) $a$ = 1.5, (d) $a$ = 2.4. The dotted curve is for $W_o$, 
solid curve is for $\tilde{W}_p$, and dotdashed curve is for 
$\tilde{W}_u$.}
\end{figure}
%
\subsection{Efficiency}
Efficiency is defined as the ratio $\eta = W/Q_{\rm h}$.
So in order to estimate efficiency, we have to estimate the amount of heat
exchanged with the hot reservoir, $Q_{\rm h}$.
However, as pointed out in the previous subsection, owing to a
complete ignorance about the association of temperature 
labels with their reservoirs, the quantity 
$Q_{\rm h}$ can be written in two ways. 
In terms of temperature $T_2$, we can either write
\be
Q_{\rm h}(T_2) = \frac{Na^2}{4}\left[\frac{1}{T_2}-1\right].
\label{heat1}
\ee
or, as
\be
Q_{\rm h}'(T_2) = \frac{Na^2}{4}\left[{\sqrt{1+\frac{1}{\theta^2}-
\frac{1}{{T_2}^2}}}-1\right].
\label{heat2}
\ee
Then the estimate for heat absorbed from the hot reservoir can be 
given by either $Q_{\rm h}(\overline{T}_2)$ or $Q_{\rm h}'(\overline{T}_2)$.

It follows that the efficiency can be 
estimated in two ways: $\tilde{\eta}_1=\tilde{W}/{Q}_{\rm h}(\overline{T}_2)$ 
or $\tilde{\eta}_2=\tilde{W}/{Q}_{\rm h}'(\overline{T}_2)$,
where $\tilde{W}$ is given by Eq.(\ref{w2es}).
Explicitly, we obtain
\bea
\tilde{\eta}_{1} = \frac{\sqrt{3\theta^2-2\theta+3}-(1+\theta)}{1-\theta},
\eea
and
\bea
\tilde{\eta}_{2} = \frac{\sqrt{3\theta^2-2\theta+3}-(1+\theta)}
{\sqrt{3\theta^2-2\theta+3}-2\theta}.
\eea
We now compare the above estimates with the efficiency
at optimal work, which by using Eq. (\ref{topt}), is given as: 
\bea
{\eta}_o=2\left(\frac{\sqrt{2(1+\theta^2)}-(1+\theta)}{\sqrt{\frac{1+\theta^2}{2}}
-\theta}\right).
\eea
Fig. 3 shows these comparative plots.  
The estimates expanded near equilibrium are as follows:
\bea
\tilde{\eta}_{1} & \approx&  \frac{\eta_c}{2} + \frac{{\eta_c}^2}{4} + 
\frac{1}{16}{\eta_c}^3 + O[{\eta_c}^4],\\
\tilde{\eta}_{2} & \approx&  \frac{\eta_c}{2} - \frac{1}{16}{\eta_c}^3 + O[{\eta_c}^4],
\eea 
\par\noindent 
whereas near equilibrium,  the efficiency at optimal work, behaves as:
\bea
{\eta}_o & \approx & \frac{\eta_c}{2} + \frac{{\eta_c}^2}{8} + O[{\eta_c}^4].
\eea
In this situation, we observe agreement with the optimal behavior 
only upto first order. 
\begin{figure}
\includegraphics[width=2.5in]{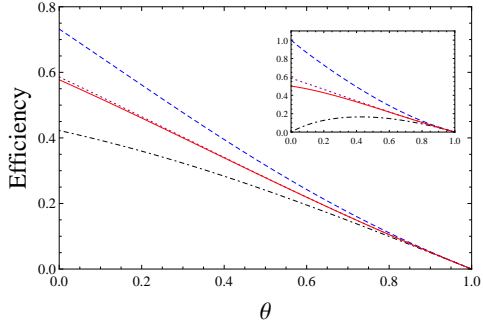} 
\caption{Efficiency vs. $\theta$. The 
solid curve is the mean estimate $\tilde{\eta}$, closely following the 
dotted curve which is for optimal value $\eta_o$. The top and
the bottom curves are $\tilde{\eta}_{1}$ and $\tilde{\eta}_{2}$. 
The inset shows the corresponding quantities with the use of uniform prior.}
\end{figure}
On the other hand, if we define a mean estimate for efficiency as 
$\tilde{\eta}=(\tilde{\eta}_{1}+\tilde{\eta}_{2})/2$,
then the agreement of this mean with the optimal behavior is upto third order.
The use of a mean estimate can be justified as follows. 
We have two hypotheses, whether the heat extracted is given by 
Eq. (\ref{heat1}) or (\ref{heat2}). According to Laplace's principle of insufficient
reason \cite{Laplace1774}, when we do not have specific reason to prefer 
one hypothesis over another, then we should assign
equal weights to the inferences following from each 
of these hypotheses. In our case, we have assumed
complete ignorance about the labels attached with 
final temperatures and so each expression for 
$Q_{\rm h}$ above is equally valid.
In this sense, it is reasonable that the most unbiased estimate    
be based on an equally-weighted mean of the different estimates.

It is to be noted that this property also emerges during
the use of a uniform prior. 
 Thus we can see analytically that in the near equilibrium case for small $a$ values, 
the uniform prior as well as the non-uniform  prior both replicate the
optimal properties of the work as well as efficiency to terms beyond linear
response.  However, as corroborated by the full numerical calculations for
arbitrary $a$ and  for general temperature differences (Figs. 2 and 3),  
it is apparent that the estimates with the 
non-uniform prior provide a quite good agreement with the optimal
properties than the uniform prior. 
%
%
\section{Thermal Interaction}{\label{th}}
In this section, we study another well-known process in which 
two finite reservoirs interact thermally 
with each other, while conserving the total energy. A little amount of heat energy
is quasi-statically removed from the hot reservoir and deposited in the same manner
with the cold reservoir. The optimal process is the one which terminates at a common
temperature. As is known,
there is a net entropy production in the reservoirs.  We consider a 
situation in which the final state
after interaction is not specified, and so we have to estimate the 
final state. As in previous sections, we need to derive the prior 
to quantify our uncertainty about the 
value of the final temperatures.

 Now the prior  $\pi(T)$ will be derived from the information that the 
 transfer of an infinitesimal
 energy from the hot to the cold reservoir, does not change the total energy: 
\be
dU_1 + dU_2 = 0, 
\ee
which can be written as:
\bea
\frac{\partial U_1}{\partial T_1} dT_1 +
\frac{\partial U_2}{\partial T_2} dT_2 & = & 0.
\eea
This yields
$|{dT_1}/dT_2| =  C_2(T_2)/C_1(T_1)$.
Again similar to Eq. (\ref{pr}), we have $|{dT_1}/dT_2| = \pi (T_2)/ \pi(T_1)$,
Identifying these two conditions and using separation of variables, we   
obtain the prior for each of final temperatures, in the form 
\be
\pi(T_i) = \frac{C_i(T_i)}{\int_{\theta}^{1} C_i(T_i)dT_i}.
\label{prior2}
\ee
The estimation of various quantities follows the similar procedure as highlighted
in Section \ref{m}. 
Here also, the constraint equation of energy conservation: $U_1 + U_2 = U_+ + U_-$, cannot
be solved to yield an explicit relation between $T_1$ and $T_2$. We can perform
analytical calculations only for the high temperatures case.
The main quantity of interest will be the net entropy production.
\subsection{High-temperature limit}
The explicit relation between $T_1$ and $T_2$ is obtained by using Eq. (\ref{int2}) 
and total energy conservation of the reservoirs. This yields 
\be
T_1  =  \left(1+\frac{1}{\theta}-\frac{1}{T_2}\right)^{-1}.
\label{rel2}
\ee
For the optimal process, $T_1=T_2=T_c$, where $T_c = 2 \theta /(1+\theta)$.
The expected value of one of the temperatures ($T_i$) over the informative prior 
(Eq. (\ref{prior2})), is calculated as:
\be
\overline{T}_i = \frac{\theta \ln({1/\theta})}{(1-\theta)}.
\ee
The entropy produced $\bigtriangleup S = S_1 + S_2 - S_+ - S_- $, 
can be written using Eq. (\ref{ent2}) as:
\be
\bigtriangleup S = \frac{N a^2}{8}\left(1+\frac{1}{\theta^2}-
\frac{1}{{T_1}^2}-\frac{1}{{T_2}^2}\right),
\ee
which can be expressed as function of one variable, using Eq. (\ref{rel2}).
Then the estimate for entropy production 
in the high-temperature limit, is given by replacing $T_i$ with $\overline{T_i}$.
The estimation was done with informative as well as uniform prior and
compared with the optimal behavior. Fig. 4 shows the comparison 
in the limit $a/T \ll 1$.
\begin{figure}
\includegraphics[width=2.5in]{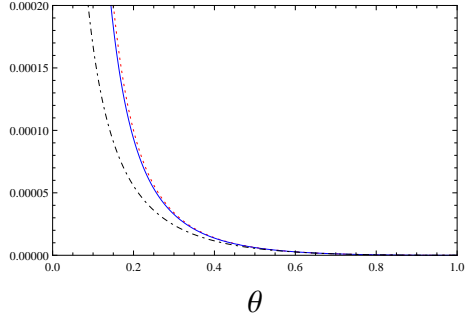} 
\caption{Entropy production, scaled by $N$, 
as a function of $\theta$ for $a=0.01$ in  thermal contact process.
The top, dotted curve is the optimal entropy production, 
the middle, solid curve is the estimate using informative
prior, and lower, dotdashed curve is from the use of uniform prior.}
\end{figure}
When we expand the estimates for entropy production in near-equilibrium regime, we get:
\bea
{\bigtriangleup {S}}_p \approx {\bigtriangleup {S}}_u &=& \frac{N a^2}{16}(1-\theta)^2
        + \frac{N a^2}{8}(1-\theta)^3+ O[1-\theta]^4. \nonumber \\
\label{entseries}
\eea
These estimates show agreement with the optimal entropy production  
upto third order, as in the case of estimated work in Section \ref{highT}. 

Finally, for general $a$ values, the numerical calculations show that 
the estimated entropy production  with the derived prior  shows a good agreement
with the optimal entropy production, along similar lines as for estimates of extracted work. 
\section{Discussion}
In this section, we try to gain an  
insight into the form of prior. First we note that 
the prior for final temperature in the entropy conserving process, Eq. (\ref{pr}), 
can be reexpressed as:
\be
P(T) dT = \frac{dS}{(S_+-S_-)}.
\ee
Thus the derived prior is equivalent to 
 a uniform prior in terms of the entropy, defined over the interval $[S_-,S_+]$.
Similarly, the prior for the energy conserving process, Eq. (\ref{prior2}), implies
 a uniform prior over the energy of a reservoir: $\pi (T) dT = dU/(U_+-U_-)$.
Thus our particular choice of prior for temperature, implies a uniform prior 
density for the quantity being conserved in the process. 

Secondly, the proposed prior  lends a specific meaning to the 
final common temperature ($T_c$). For the optimal entropy 
conserving process, 
the change in entropy of a reservoir is given by: $S_+ - S_c = S_c - S_-$. 
This can be written in integral form as:
\be
\int_{S_c}^{S_+} dS  = \int_{S_-}^{S_c} dS.
\ee
As the prior density is uniform in terms of entropy, so
we can write 
\be
\int_{S_c}^{S_+} p(S) dS  = \int_{S_-}^{S_c} p(S) dS,
\ee
where $p(S) = 1/(S_+-S_-)$. Thus our choice of prior
implies that we are assigning {\it equal} probability
(one-half each) that 
entropy $S$ of a reservoir may lie in 
the interval $[S_-,S_c]$, or in the interval $[S_c,S_+]$.
A similar statement can be made in terms of $T_c$.
Thus $T_c$ is the median of  prior $P(T)$, 
on either side of which 
we expect equal chances that the final temperature may lie.

The whole analysis can also be looked at in terms 
of macrostates. 
Consider the basic question in equilibrium thermodynamics 
\cite{Callen1985}. Given that entropy
is conserved for a bipartite system whose total energy is allowed
to vary, what is the most likely state of the system?
The answer is given as the equilibrium state which has 
 minimum total energy for the given value of the total entropy. 
(In terms of work, it translates into an extraction of maximum work.)
The agreement of our estimates with
the optimal work and the corresponding efficiency which was found in Section IV, shows that 
we are able to estimate the equilibrium state consistent with
the constraints, without explicitly doing an optimization.  

 Finally, let us analyse 
 the expected value of temperature as defined by $\overline{T} = \int T P(T)\,dT$.
 For the entropy conserving process, by using Eq. (\ref{pr}), 
 the estimate for temperature has the general
 form:
 \bea
 \overline{T} &=& \frac{1}{N} \int_{T_-}^{T_+} {C(T)}\;dT \nonumber\\
              &=&   \frac{1}{N} \int_{U_-}^{U_+} \! dU \nonumber \\
              &=& \frac{(U_+-U_-)}{(S_+ - S_-)},
              \label{tmean}
 \eea
 where $N = \int C/T \; dT $. This suggests that $\overline{T}$ is the 
 estimate for the derivative
 of the function $U(S)$ whose values at two points, 
 $U_+(S_+)$ and $U_-(S_-)$, have been given. We note that the above intuitive meaning 
 arises naturally within the energy representation \cite{Callen1985}.
 Similarly, if we consider pure thermal interaction, the prior is given as:
 $\pi(T) dT= C dT/\int C dT $. While there is no simple
 interpretation for the expected value of $T$ in this case, however the expected
 value of the inverse temperature $\beta = 1/T$, is given simply as
 \be
 \overline{\beta} =  \frac{(S_+-S_-)}{(U_+ - U_-)}.
 \ee
 So here, $\overline{\beta}$ can be regarded as an estimate
 for the derivative of the function $S(U)$, when its values
 at two points, $S_+(U_+)$ and $S_-(U_-)$, have been given.
 This is also consistent with the fact that  
 in the entropy represntation, $\beta$ is the derivative of
 the function $S(U)$. 

 Concluding, we have extended the basic approach suggested 
 in \cite{PRD2013}, to quantify uncertainty as subjective
 probability, in constrained thermodynamic processes. 
 In this paper, we considered entropy conserving as well
 as energy conserving processes. We argued for and proposed
 a general prior while incorporating the prior information
 about the process. Depending on the kind of process, 
 this general form is applied to the case of
 two spin-1/2 systems as heat reservoirs, to 
  estimate the maximum extracted work and the corresponding
 efficiency or the net entropy production. 
 The agreement
 with optimal values of these quantities are shown 
 in high temperatures limit, as well as by numeric calculations.
 Finally, in order to elucidate the meaning 
 of the prior, we found certain points of consistency 
 of our approach with 
 the standard axiomatic thermodynamic framework.
 In our opinion, our approach seems applicable to 
 quantify uncertainty in a subjective sense, 
 for other constrained optimization problems.
 Further, some interesting future problems 
 may be cited as: generalization to 
 multipartite systems and the use of non-identical reservoirs.
\section{Acknowledgements}
RSJ acknowledges financial support from the Department of
Science and Technology, India under the research project
No. SR/S2/CMP-0047/2010(G). PA is thankful to University
Grants Commission, India for Research Fellowship.


\begin{thebibliography}{50}
\bibitem{Comment_random} For simplicity, we denote both the uncertain
variable as well as its value by a capital symbol.
\bibitem{Comment_uniform} As an alternative,  
we may choose to ignore the mutual dependence of parameters $T_1$ and $T_2$
while assigning the prior. 
Then a plausible prior is the uniform density, $P(T_i) = {\rm constant}$.
\end{thebibliography}
\end{document}